\documentclass[pre,amsmath,twocolumn]{revtex4}
\usepackage{graphicx} 
\usepackage{epsfig}
\usepackage{amssymb}

\newcommand{\be}{\begin{equation}}
\newcommand{\ee}{\end{equation}}
\newcommand{\lb}{\ell_B}
\newcommand{\eff}{\hbox{\scriptsize eff}}
\newcommand{\bare}{\hbox{\scriptsize bare}}
\newcommand{\sat}{\hbox{\scriptsize sat}}

\newcommand{\Zeff}{Z_{\hbox{\scriptsize eff}}}
\newcommand{\Zbare}{Z_{\hbox{\scriptsize bare}}}
\newcommand{\Zsat}{Z_{\hbox{\scriptsize sat}}}

\begin{document}
\title{The interplay between screening properties and colloid
anisotropy: towards a reliable pair potential for disc-like charged 
particles
}
\author{Rafael Agra$^1$, Emmanuel Trizac$^1$ and Lyd\'eric Bocquet$^2$}

\affiliation{$^1$ Laboratoire de Physique Th\'eorique, Universit\'e
de Paris XI, B\^atiment 210, 91405 Orsay C\'edex, France}

\affiliation{$^2$ Laboratoire de Physique de la Mati\`ere Condens\'ee et 
Nanostructures, Universit\'e Lyon 1, 69622 Villeurbanne, France}


\date{\today}

\begin{abstract}
The electrostatic potential of a highly charged disc 
(clay platelet) in an electrolyte is investigated in detail. 
The corresponding non-linear Poisson-Boltzmann (PB) equation is
solved numerically, and we show that the far-field
behaviour (relevant for colloidal interactions in dilute 
suspensions) is exactly that obtained within linearized 
PB theory, with the surface boundary condition of a uniform
potential. The latter linear problem is solved by a new semi-analytical
procedure and both the potential amplitude (quantified by 
an effective charge) and potential anisotropy coincide
closely within PB and linearized PB, provided the disc 
bare charge is high enough. This anisotropy
remains at all scales; it is
encoded in a function that may vary over several orders
of magnitude depending on the azimuthal angle under which
the disc is seen. The results allow to construct a pair
potential for discs interaction, that is strongly 
orientation dependent.
\end{abstract}

\maketitle
\section{Introduction}

Clays, in the generic form of charged platelets, enjoy widespread use
in applications ranging from drilling, rheology modification (for paints,
cosmetics, cleansers...), catalysis etc. As a significant component
of soils, clays are also of importance for crop production. The
difficulty of synthesizing clays with well controlled properties
(size, composition, charge...) has long hindered their fundamental
study. The situation has considerably changed in the last ten years,
with the increasing availability of customized synthetic clays,
among which Laponite is a prominent example. Yet, our understanding
of such systems is rudimentary (see e.g. 
\cite{Mourchid,Bonn,Nicolai,JPCM,Levitz,Michot,Rowan,Beek} 
and references therein).

A reasonable model for Laponite platelets is that of uniformly 
charged and infinitely thin discs \cite{Rque1}.
In this paper, the focus will be on electrostatic interactions between
identical charged discs, a crucial ingredient for understanding 
the phase behaviour and stability of clays in suspensions.
The high anisotropy of these objects makes analytical progress
difficult. In addition, these discs are typically highly charged,
and the electrostatic coupling with their electrolytic environment
(microscopic charged species) needs to be described by non-linear
theories: the plain linear Debye-H\"uckel approach should fail. We 
will work here in the common framework of non-linear Poisson-Boltzmann
(PB) theory, where the (dimensionless) 
electrostatic potential outside the charged 
macro-ions obeys an equation of the form $\nabla^2 \phi =\kappa^2 \sinh\phi$,
assuming for simplicity monovalent microions only, the density of which
governs the screening length $\kappa^{-1}$ (the Debye length).

In a solution, the typical distance between macroions is often larger than
the Debye length (this condition requires a minimal but nevertheless
small amount of salt). At these ``large'' scales, the potential created by 
a given disc is small enough --compared to thermal agitation--
to allow for the linearization of PB equation: 
$\nabla^2 \phi \simeq \kappa^2 \phi$. Accordingly, the potentials
within non-linear PB on the one hand, and linearized PB theory 
with {\em a suitably chosen boundary condition} on the other hand,
coincide at large enough distance from the colloids, be they of
discotic or other shapes. An analytical treatment within linearized
PB (LPB) is of course considerably simpler than within PB, but the
above remark may be of little practical help if one is not able to
derive the relevant boundary condition on the colloid
(effective potential), such that 
the corresponding LPB solution reproduces the PB one in the region of low enough
potential. Close to the colloids, non-linear effects prevail 
(LPB and PB solutions strongly differ), and
broadly speaking, microions --essentially counterions-- suffer there a
high electrostatic coupling and may be considered as ``bound''. They 
decrease the bare charge of the colloid so that its electrostatic
signature at large distance defines an effective charge which is usually
smaller in absolute value than the bare one (close to the colloid, 
the effective potential
is accordingly smaller than its non-linear counterpart). 

For a unique charged sphere in an electrolyte, PB and LPB theories give
rise to the same far-field behaviour, of Yukawa type [$\exp(-\kappa r)/r$,
where is $r$ is the radial coordinate];
non-linear effects only affect the prefactor
(from which the effective charge is defined), preserving the functional
form of the potential \cite{Belloni,Hansen,Levin}. 
The same remark equally applies to an
infinite rod. The situation changes, however, for anisotropic 
objects such as discs or finite size rods \cite{Chapot},
where non-linear screening phenomena generically  
affect the functional form of the potential and cannot be subsumed in
an effective scalar quantity (effective potential or charge). 
In other words, whereas the symmetry of the effective colloid
clearly remains spherical in the case of spheres, predicting the
symmetry of the effective charge distribution and associated
electrostatic potential for a highly charged
disc is a non trivial question.

It is the purpose of the present work to study how non-linear screening effects 
and anisotropy conspire to affect the far-field behaviour in the case of 
discs. It has been shown in \cite{Presc} that highly charged spheres
and infinite rods may be considered as objects of constant effective
potential $\phi_{\eff}$ (in the sense that $\phi_{\eff}$ becomes independent
of physico-chemical parameters, provided that $\kappa a>1$ where 
$a$ is the colloid radius. The complementary results reported here
indicate that the constant potential picture goes in fact beyond this analysis, 
and give the correct symmetry of the effective charge distribution 
onto the disc. Such a boundary condition (within LPB theory) 
produces the same electrostatic potential as a highly charged disc within PB.
A physical argument allowing to anticipate this correspondence will 
be presented in section \ref{sec:why}. Since the exact LPB solution for a disc
held at constant potential in an electrolyte is not known, 
we will introduce in section \ref{sec:fredholm} 
a semi-analytical procedure to solve this problem. 
The characteristic features of the electrostatic potential relevant
for clay discs will be obtained, and in order to assess the validity of
the constant potential picture, the corresponding electrostatic potential
will be compared in section \ref{sec:pb} to the solutions of the full
non-linear PB theory. The latter will be obtained 
through an iterative numerical procedure. From these results,
a pair potential will be constructed for charged discs, that has the
same status as the celebrated Derjaguin-Landau-Verwey-Overbeek expression
relevant for spheres \cite{Belloni,Hansen,Levin}, and which includes
charge renormalization.
Concluding remarks will finally be presented in sections \ref{sec:remarks}
and \ref{sec:concl}.

\section{The constant effective potential picture: why?}
\label{sec:why}

A disc of radius $a$ and uniform surface charge density
$\sigma_{\bare}=Z_{\bare}\,e/(\pi a^2)$ 
is immersed in an infinite sea of electrolyte with bulk density
$n_s$. From the permittivity $\varepsilon$ of the solvent,
the Bjerrum length is defined as $\lb = e^2/(\varepsilon kT)$,
where $kT$ is the thermal energy and $e$ denotes the elementary charge. 
Within non-linear Poisson-Boltzmann
theory, the dimensionless potential [$\phi=e\varphi/(kT)$,
$\varphi$ being the original electrostatic potential]
created by the disc obeys the following
equation
\be
\nabla^2 \phi = \kappa^2 \sinh(\phi)
\label{eq:pb}
\ee
where $\kappa$ is the inverse Debye length defined through
$\kappa^2 = 8 \pi \lb n_s$. For convenience, $\phi$ is chosen
to vanish at infinity. Eq. (\ref{eq:pb}) holds outside the disc.

\begin{figure}[htb]
$$\epsfig{figure=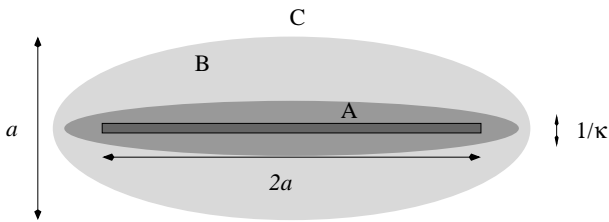,width=8.cm,angle=0}$$
\caption{\label{fig:fig1} Schematic side view of the charged
platelet}
\end{figure}
Considering a highly charged disc with furthermore $\kappa a >1$,
one may partition space into three regions, as sketched  
in Fig. \ref{fig:fig1}.
In region A, the electrostatic coupling between the colloidal
disc and the microions is most important and one has $\phi>1$.
Outside A, in regions B and C, one has $\phi<1$ and PB
equation (\ref{eq:pb}) may be linearized: the corresponding Helmholtz-like
LPB equation reads 
\be
\nabla^2\phi=\kappa^2\phi.
\label{eq:lpb}
\ee
In addition, in region B,
$\phi$ is of unidimensional character and well approximated
by the potential created by an infinite charged plane. {\it A contrario},
in region C, $\phi$ regains its full 3D nature (2D here with the
present azimuthal symmetry). The lateral extension of the ``non-linear region''
A is given by $\kappa^{-1}$ while $a$ measures the extension
of region B. Since we assume $\kappa a>1$, we have A$\,\subset\,$B and
moving  away from the disc,  non-linear effects disappear
before the finite size of the disc becomes relevant. 
In other words, B is the
non vanishing intersection between
the ``linear'' region 
and its one-dimensional counterpart where the
solution of Eq. (\ref{eq:pb}) takes the form \cite{Andelman}
\begin{eqnarray}
\phi_{\text{1\sc D}} &=& 4\, \hbox{arctanh}\left(\gamma e^{-\kappa z}\right)\\
&\simeq& 4 \gamma e^{-\kappa z} \quad\hbox{in region B.}
\label{eq:tourdefrance}
\end{eqnarray}
In these expressions, $z$ denotes the distance to the plane
and, assuming without loss of generality 
a positive bare charge $\sigma_{\bare}$, 
$\gamma$ is the positive root of the quadratic equation
\be
\gamma^2 - \frac{\kappa e}{\pi \sigma_{\hbox{\tiny bare}} \lb} \,\gamma+1  =0.
\label{eq:gamma}
\ee
In colloidal dispersions, the relevant range for the interactions is that of
far-field (except for dense systems) and the behaviour in the non-linear
region (A) is of little interest. From Eq. (\ref{eq:tourdefrance}), 
it appears that the potential 
felt in the linear region B+C, when extrapolated to contact ($z=0$), reads
$\phi_{\eff} = 4 \gamma$. As a consequence, solving LPB equation
(\ref{eq:lpb}) with this boundary condition should provide the
same potential outside region A as the PB solution of equation 
(\ref{eq:pb}). 

The above remarks follow from the constraint $a > \kappa^{-1}$ 
and apply irrespective of the value of the bare charge. In particular,
$\sigma_{\bare}$ (hence $\phi_{\eff}$ through $\gamma$) 
may be position dependent on the disc. However, we are interested here
in highly charged discs for which the non-linear region A exists
(for low $\sigma_{\bare}$, region A disappears; PB and LPB solutions
coincide {\em at all distances} and the issue of effective potentials
becomes trivial: effective and bare charges are equal).
From Eq. (\ref{eq:gamma}), it appears that $0<\gamma<1$
and that
$\gamma \to 1$ when 
$\sigma_{\bare}$ becomes large, so that $\phi_{\eff} \to 4$,
and the field created is independent of the bare charge.
More details concerning the phenomenon
of effective charge saturation may be found in \cite{Presc}.

We conclude here that a highly charged disc should effectively 
behave as a constant potential object treated within a linear theory. 
The corresponding LPB
problem will be addressed in the following section but we emphasize
before that the argument developed here provides
$\phi_{\eff}$ to leading order in $\kappa a$. On the basis of
the behaviour of charged spheres for which the curvature correction 
has been computed in \cite{JPA} [leading to the
result $\phi_{\eff} = 4 + 2/(1+\kappa a)+{\cal O}(\kappa a)^{-2}$], 
we anticipate that 
$\phi_{\eff}$ may exceed the threshold 4. A similar behaviour
is observed for cylinders of infinite length \cite{JPA}.

\section{Semi-analytical solution of the Dirichlet linearized PB problem}
\label{sec:fredholm}

\subsection{Methodology}
\label{ssec:3a}
If the solution of LPB equation (\ref{eq:lpb}) with Dirichlet
boundary condition $\phi=\phi_0$ was known analytically, 
the effective charge of highly charged discs \cite{Rque2} would follow 
immediately, enforcing $\phi_0=4$ on the disc surface. 
Unfortunately, such a solution only exists in
vacuum (i.e. when $\kappa=0$ \cite{Jackson}). To our knowledge,
the only solution known at finite $\kappa$ is that associated to
Neumann boundary conditions (uniform surface charge) \cite{PRE97}.
To solve the Dirichlet problem, we have therefore developed a 
semi-analytical procedure where the problem at hand is recast 
into a Fredholm integral equation (see below and appendix \ref{app:A}).

The general solution of Eq. (\ref{eq:lpb}) may be written
assuming both cylindrical symmetry around an axis $(Oz)$,
and reflection symmetry $z \leftrightarrow -z$:
\be
\phi(\rho,z) = \int_0^\infty A(k) \,J_0(k\rho)\, e^{-\sqrt{k^2+\kappa^2}|z|} \, dk.
\label{eq:sollpb}
\ee
In this relation, $(\rho,z)$ denote the cylindrical coordinates and
$J_0$ is the Bessel function of order 0. The difficulty in the
present situation is that the boundary conditions imply 
that the weight function $A(k)$ obeys the mixed system
\begin{eqnarray}
&&\int_0^\infty A(k) \,J_0(k\rho)\, dk=\phi_0 \quad \hbox{for }\rho<a
\label{eq:fa}\\
&&\int_0^\infty \sqrt{k^2+\kappa^2}\, A(k) \,J_0(k\rho) 
\, dk=0 \quad \hbox{for }\rho>a
\label{eq:fb}
\end{eqnarray}
where the second equation follows from the vanishing of the normal electric
field $\partial_z \phi$ on the symmetry plane $z=0$.
Starting from Eqs. (\ref{eq:fa}) and (\ref{eq:fb}), the problem 
is rephrased in terms of an integral equation, solved numerically,
from which the function $A(k)$ is computed (see Appendix \ref{app:A}).
The potential then follows from (\ref{eq:sollpb}).

\subsection{Properties of the solutions}

\begin{figure}[tbh]
\vskip 2mm
$$\epsfig{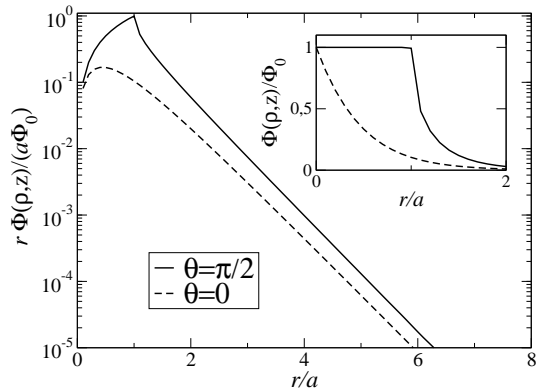}$$
\caption{\label{fig:fig2} Solution of LPB equation (\ref{eq:lpb}) 
with Dirichlet boundary condition of a uniform potential 
$\phi=\phi_0$ on the disc surface.
The quantity $r\phi$ is shown on a linear-log scale
to emphasize the far-field behaviour. Here, $\kappa a=2$ and
$r=\sqrt{\rho^2+z^2}$ denotes the distance to the disc center.
The continuous curve shows the potential in the direction $\theta=\pi/2$
(as a function of $r/a=\rho/a$),
whereas the dashed line shows the behaviour as a function of $r/a = z/a$
along normal axis $\rho=0$ ($\theta=0$).
The inset shows $\phi/\phi_0$ on a linear scale,
again in the two perpendicular directions $\theta=0$ and $\theta=\pi/2$.}
\end{figure}

A typical solution is shown in Fig. \ref{fig:fig2}. In the remainder,
the variable $\theta \in [0,\pi/2]$ denotes the angle between a given direction
and the normal to the disc ($\theta=\pi/2$ in the symmetry plane $z=0$
and $\theta=0$ along the normal to the disc, i.e. when $\rho=0$).
It may be observed that the potential is anisotropic at all
distances, a generic feature of screened electrostatics \cite{JPCM,Chapot}:
the behaviours for $\theta=0$ and $\theta=\pi/2$ strongly differ,
at all scales.
The anisotropy of the potential at large scales is encoded
in a function $f(\kappa a,\theta)$, such that expression (\ref{eq:sollpb})
may be written, for $\kappa r \gg 1$,
\be
\phi(r,\theta) \,\sim \, Z\, \lb\, f(\kappa a,\theta) \, \frac{e^{-\kappa r}
}{r}
\,+\, {\cal O}\left(\frac{e^{-\kappa r}}{r^2}\right), 
\label{eq:ff}
\ee
where $r=(\rho^2+z^2)^{1/2}$ again denotes the distance to the disc center.
In Eq. (\ref{eq:ff}), the total charge $Z$ of the platelet appears.
$Z$ is the integral over the disc surface of the surface charge
density $\sigma(\rho)$ [$Ze=\int_{\hbox{\tiny disc}} \sigma({\bf s})d^2 {\bf s}$].
This density turns out to be related to the anisotropy
function through 
\be
f(\kappa a,\theta) = \int_{\hbox{\scriptsize disc}} 
\frac{\sigma({\bf s})}{Z e}\, \exp\left(-\kappa\, {\bf \hat r}\cdot{\bf s}\right)
\, d^2{\bf s},
\label{eq:fsigma}
\ee
${\bf \hat r}$ being a unit vector pointing
in the $\theta$ direction. As expected, without electrolyte,
$\kappa$ vanishes so that $f=1$ and the potential in (\ref{eq:ff}) takes the familiar
form of an isotropic Yukawa expression.

The anisotropy function $f$ and the total charge $Z$,
are the key quantities governing far-field behaviour. 
Note that $Z$ is not known a priori since only the
surface potential is imposed. It may be shown 
that $f$ is related to the weight function $A(k)$ appearing
in (\ref{eq:sollpb}) through
\be
\frac{Z\,e}{\varepsilon} f(\kappa a,\theta) \,=\,
 -i\,\frac{A(i\,\kappa  \sin\theta)}{\tan\theta}.
\ee
For small arguments, one has $A(x)\propto x $ so that $f(\kappa a,0)=1$,
which also means that the total charge is directly accessible through the
behaviour along the $\theta=0$ axis~:  
\be
\phi(r,0) \sim  Z\, \lb\, \frac{e^{-\kappa r}
}{r}.
\label{eq:normalaxis}
\ee
Note also that $f(\kappa a,0)=1$ directly follows from 
(\ref{eq:fsigma}) since ${\bf \hat r}\cdot{\bf s}=0$ when $\theta=0$.

\begin{figure}[tbh]
\vskip 2mm
$$\epsfig{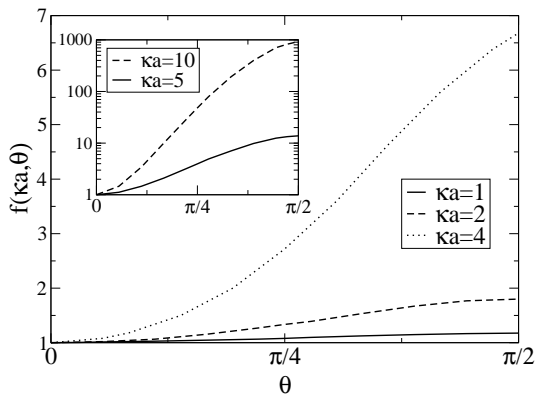}$$
\caption{\label{fig:fig3} Anisotropy function for $\kappa a$=1, 2 and 4
as a function of azimuthal angle. The inset shows the results on a linear-log
scale, for $\kappa a=5$ and $\kappa a=10$.}
\end{figure}
\begin{figure}[tbh]
\vskip 2mm
$$\epsfig{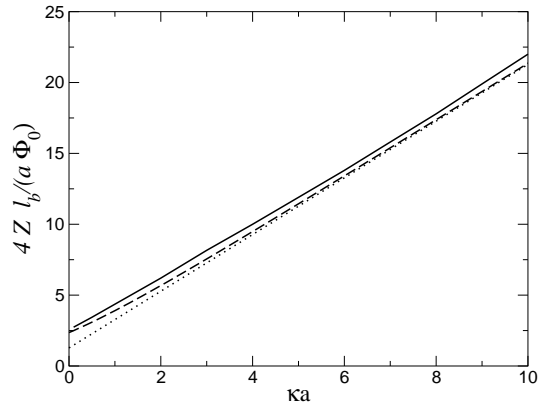}$$
\caption{\label{fig:fig5} Charge $ 4 Z \lb/(a \phi_0)$ as a function of
$\kappa a$ (continuous curve). The dashed line shows the result obtained
within the simplified two parameter model [Eqs. (\ref{eq:sig2param})
and (\ref{eq:systa})-(\ref{eq:systb}), see text]. The dotted line 
corresponds to Eq. (\ref{eq:Zestime}). }
\end{figure}

Figure \ref{fig:fig3} shows $f(\kappa a,\theta)$ for different 
salinity conditions. This function increases with $\theta$
and may take large values when $\kappa a$ exceeds a few units
(see the inset where the $y$-axis is shown in log scale).
On the other hand, for $\kappa a<1$, $f$ remains close to unity 
in all directions.
The potential is strongest in the disc plane ($\theta=\pi/2$), and increasing
screening ($\kappa$), one also strongly increases the anisotropy
of the electrostatic potential. For a reasonable value 
$\kappa a=10$, $f(\theta)$ varies by almost 3 orders 
of magnitude (a factor 930). In figure \ref{fig:fig5},
the complementary information concerning the total charge
$Z$ is displayed. This quantity will be further discussed
in section \ref{ssec:approxf}. Note that $f(\kappa a,\theta)$ does
not depend on $\phi_0$, since it probes the repartition 
of surface charge distribution, not its overall magnitude.
On the other hand, $Z$ scales linearly with $\phi_0$.

\subsection{An approximate expression for the anisotropy function
and charge}
\label{ssec:approxf}

It is instructive and useful for practical purposes
to have an approximate analytical expression for the function 
$f(\kappa a,\theta)$. From Eq. (\ref{eq:fsigma}), that may
rewritten
\be
f(\kappa a,\theta) \,=\, 
\frac{2 \pi}{Z e} \int_0^a I_0(\kappa \rho \sin\theta)\, \sigma(\rho)\, 
\rho \, d\rho,
\label{eq:fI0}
\ee
this amounts to look for an approximate expression for the surface charge
density $\sigma$. To this aim, we recall \cite{Landau} 
that for an ideal conducting disc when
$\kappa =0$, $\sigma$ diverges in the vicinity of the edge ($\rho\to a$)
since 
\be
\sigma(\rho) \,=\, \frac{e\,\phi_0}{2 \pi^2 a \,\lb} \, \frac{1}{\sqrt{1-(\rho/a)^2}}.
\label{eq:sigmadiverge}
\ee
Recall that $\phi_0$ denotes the dimensionless electrostatic potential
$\phi = \varphi /(\varepsilon k T)$.
The singularity of the electric field near sharp edges,
which is the reason for the efficiency of lightning conductors, also
pertains in presence of an electrolyte. We indeed show in appendix
\ref{app:C} that when $\kappa\neq 0$, the Dirichlet solution to Eq.
(\ref{eq:lpb}) exhibits a similar divergence as that present 
in Eq. (\ref{eq:sigmadiverge}),
namely 
$\sigma \propto (a-\rho)^{-1/2}$. 

\begin{figure}[tbh]
\vskip 2mm
$$\epsfig{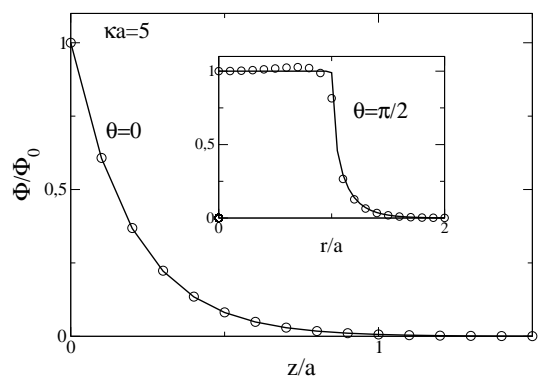}$$
\caption{\label{fig:fig6} Electrostatic potential following from the
two parameters approximation (circles) 
compared to the exact result (continuous curve,
obtained following the method
detailed in section \ref{ssec:3a} and Appendix \ref{app:A}). 
The main graph corresponds to $\theta=0$ and the inset to $\theta=\pi/2$.}
\end{figure}

\begin{figure}[tbh]
\vskip 2mm
$$\epsfig{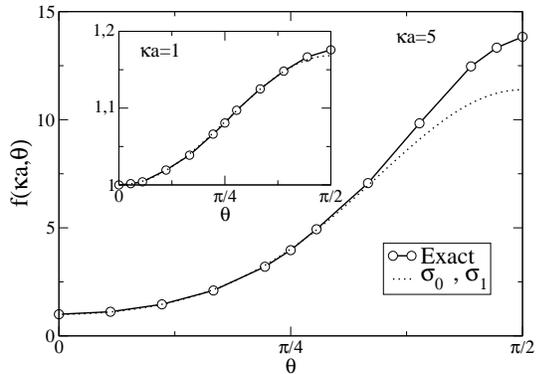}$$
\caption{\label{fig:fig7} Anisotropy function resulting from 
the two parameters approximation (\ref{eq:sig2param}) where 
$\sigma_0$ and $\sigma_1$ are determined from 
(\ref{eq:systa})-(\ref{eq:systb}),
compared to the exact result.
The two above parameters are plotted in Fig. \ref{fig:fig8}.
Here, $\kappa a=5$ whereas $\kappa a=1$ in the inset. }
\end{figure}

\begin{figure}[tbh]
\vskip 2mm
$$\epsfig{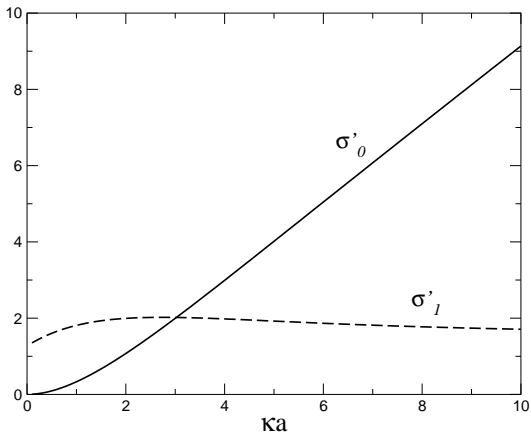}$$
\caption{\label{fig:fig8} Dimensionless charges
$\sigma'_{0,1}=2\pi a \lb\, \sigma_{0,1}/(e \phi_0)$  
following from (\ref{eq:sig2param}) supplemented with 
Eqs. (\ref{eq:systa})-(\ref{eq:systb}), as a function of salinity conditions.}
\end{figure}

A very simple two parameter
ansatz fulfilling this divergence requirement is
\be
\sigma(\rho) \,=\, \sigma_0 + \sigma_1 \frac{1}{2\,\sqrt{1-(\rho/a)^2}}.
\label{eq:sig2param}
\ee
From this expression, the anisotropy function may be computed and takes
the form
\be
f(\kappa a,\theta) \,=\, \frac{\sigma_0}{\sigma_0+\sigma_1} \, 
\frac{2 I_1(\kappa a \sin\theta)}{\kappa a\sin\theta} \,+\,
\frac{\sigma_1}{\sigma_0+\sigma_1} \, 
\frac{\sinh(\kappa a \sin\theta)}{\kappa a\sin\theta}
\label{eq:f2param}
\ee
In Eqs. (\ref{eq:fI0}) and (\ref{eq:f2param}), $I_0$ and $I_1$ denote
modified Bessel functions of the first kind, of order 0 and 1. 
Expressions (\ref{eq:sig2param}) and (\ref{eq:f2param}) are not
exact and there are several ways to choose the two parameters 
$\sigma_0$ and $\sigma_1$, that will be determined by two constraints.
The simplest possibility is to enforce $\phi(0,0)=\phi(a,0)=\phi_0$,
but it turned that the choice (hereafter adopted)
\begin{eqnarray}
&&\phi(0,0)  \,=\, \phi_0 
\label{eq:systa}\\
&&\langle \phi(\rho,0)\rangle \,=\, \phi_0
\label{eq:systb}
\end{eqnarray}
gave better results (the angular brackets denote average over
the disc surface). In the limit $\kappa \to 0$, expressions
(\ref{eq:sig2param}) and (\ref{eq:f2param}) 
become exact (with $\sigma_0=0$), and we expect that the comparison with exact
results at finite $\kappa$ will be all the better as $\kappa a$ is low.
It is indeed the case, but when $\kappa a=5$, the 
approximation is still reasonable (see Figs. \ref{fig:fig6}
and \ref{fig:fig7}).
By comparison with the exact solution Fig. \ref{fig:fig7}
shows that the
anisotropy of the potential is correctly captured.
The corresponding values of partial surface charges $\sigma_0$ and $\sigma_1$
are shown in Fig. \ref{fig:fig8}, where it may be observed that 
$\sigma_0$ vanishes at low salt, as expected. The associated 
total charge $Z$ is given by
\be
Z  \,=\, \frac{\pi a^2}{e}\,(\sigma_0+\sigma_1) \,=\, \frac{a \phi_0}{2 \lb}\,
(\sigma'_0+\sigma'_1),
\ee
where $\sigma'_1$ and $\sigma'_2$ are the quantities plotted in 
Fig. \ref{fig:fig8}.
The charge $Z$ is displayed in Fig. \ref{fig:fig5}, and compares 
favorably with its exact counterpart. In the above expression, 
however, $\sigma_0$ and $\sigma_1$ are functions of $\kappa a$ with
unknown analytical expression. As it is desirable to
have an analytical formula, we propose the following argument:
in the limit of large $\kappa a$, the disc essentially behaves 
as a an infinite plane, from which we deduce $Z\lb/a \sim \phi_0\kappa a/2$.
To estimate the next order correction [constant term $C$ in the expansion
$Z\lb/(a\phi_0) = (2\kappa a/ + C)/4$]
we may take the limit $\kappa a=0$ where the solution
is given by (\ref{eq:sigmadiverge}), which imposes
$Z\lb/a = \phi_0/\pi$. We therefore obtain  
\be
Z \,\frac{\lb}{a} \,\simeq\, \frac{\phi_0}{4} \, \left(2\,\kappa a \,+\, 
\frac{4}{\pi} 
\right).
\label{eq:Zestime}
\ee
Anticipating that the relevant values of $\phi_0$ are close to 4
(see sections \ref{sec:why} and \ref{sec:pb}),
we have factorized the ratio $\phi_0/4$ in the previous relation.
The quality of approximation (\ref{eq:Zestime}) 
is assessed in Figure \ref{fig:fig5},
which shows a good agreement.
On the other hand, extracting the correction factor $C$
from the large $\kappa a $ behaviour of the exact $Z$ 
displayed in Fig. \ref{fig:fig5} gives $C\simeq 1.88$,
to be compared with $C=4/\pi \simeq 1.27$ in Eq. (\ref{eq:Zestime}).

\section{Numerical resolution of the non-linear Poisson-Boltzmann theory}
\label{sec:pb}

In section \ref{sec:fredholm}, we have obtained the solution 
of linearized PB theory with uniform potential boundary 
condition on the disc. From the discussion developed in 
section \ref{sec:why}, we expect the properties described
(with $\phi_0=4$)
to characterize also the far-field created by a highly charged 
disc, treated within non-linear PB theory. In the following, 
we critically test this scenario.
We first present the numerical procedure used to solve
the non-linear PB problem.

\subsection{Green's function formalism and numerical method}
\label{ssec:nlpb}

Introducing explictly the charge density $q_d({\bf r})$ borne by the disc,
Eq. (\ref{eq:pb}) is rewritten
\be
\nabla^2 \phi = \kappa^2 \sinh(\phi) + 4 \pi \lb \,
\frac{q_d({\bf r})}{e}.
\label{eq:pbmod}
\ee
In the subsequent analysis, we will consider the case of a 
uniformly charged disc for which one has, in cylindrical coordinates
\be
q_d({\bf r}) \,=\, \sigma_{\bare}\, \delta(z) \Theta(a-\rho),
\ee
where $\Theta$ is the Heaviside step function and $\delta$ the
Dirac distribution.
However, it is important to emphasize that the results that 
will be derived are more general, and hold 
irrespective of the precise PB boundary condition on the disc, 
provided the bare disc charge is high enough
(phenomenon of effective potential saturation).

In view of a numerical resolution, it is convenient
to rewrite (\ref{eq:pbmod}) in the form 
\be
(\nabla^2 -\kappa_0^2) \phi \,=\, 
\kappa^2 \sinh(\phi) + 4 \pi \lb \,q_d({\bf r})/e -
\kappa_0^2 \phi,
\label{eq:pbinte}
\ee
where $\kappa_0$ is an arbitrary quantity 
that will be optimized in order to speed up the resulting
procedure (see below). Introducing the Green's function
\be
{\cal G}({\bf r},{\bf r}') \,=\,
-\frac{e^{-\kappa_0|{\bf r}-{\bf r}'|}}{4 \pi|{\bf r}-{\bf r}'|},
\ee
solution of
\be
(\nabla^2 -\kappa_0^2) \,{\cal G}({\bf r},{\bf r}')
\,= \,\delta({\bf r}-{\bf r}'),
\ee
we may recast (\ref{eq:pbinte}) into 
\be
\phi({\bf r}) =\!\int\! {\cal G}({\bf r},{\bf r}') 
\left[\kappa^2 \sinh[\phi({\bf r}')] + 4 \pi \lb \,\frac{q_d({\bf r}')}{e} -
\kappa_0^2\, \phi
\right]d{\bf r}'
\label{eq:resol}
\ee
The contribution arising from $q_d$ may be computed analytically:
\begin{eqnarray}
&&\!\!\!\!\!\!\!\!\!
\frac{4\pi l_B}{e} \int {\cal G}({\bf r},{\bf r}') q_d({\bf r}') \,d{\bf 
r}' =
\nonumber\\&&
2\frac{Z l_{b}}{a}\int_{0}^{\infty}\,J_{1}(ak)J_{0}(k\rho)
\frac{\exp(-|z|\sqrt{k^{2}+\kappa^{2}})}{\sqrt{k^{2}+\kappa^{2}}} dk \,\,
\label{eq:lpbneum}
\end{eqnarray}
and Eq. (\ref{eq:resol}) is solved iteratively. Starting 
from the initial guess $\phi_0=0$, the right hand side of (\ref{eq:resol}),
denoted $\phi_0^{\text{out}}$ is computed. This provides a new input potential
$\phi_1 = \alpha  \phi_0^{\text{out}} + (1-\alpha) \phi_0$,
which is itself inserted in the rhs of (\ref{eq:resol}) to produce
$\phi_1^{\text{out}}$ etc. The mixing parameter $\alpha$ is chosen
in the range $[10^{-2};10^{-1}]$.
Convergence $\phi_n\simeq \phi_n^{\text{out}}$
is generally achieved
for typically 50 to 200 iterations. The procedure may be 
accelerated starting not from $\phi_0=0$ but from 
the solution of LPB theory [known in the present
Neumann case, and given by Eq. (\ref{eq:lpbneum})].
In addition, it seems that the optimal choice for the numerical
screening parameter $\kappa_0$ is $\kappa_0 \simeq \kappa$.
We have checked that the solutions found were independent of
$\kappa_0$ (as they should) by changing this parameter in the range
$[\kappa/5; 5\kappa]$.
The previous procedure 
bears similarities with the one used in Ref. \cite{Raul},
where a confined geometry was considered (whereas the situation
is that of infinite dilution here).

\subsection{Results}

From the method sketched in section \ref{ssec:nlpb}, the numerical
solution of the non-linear PB equation (\ref{eq:pb}) may be
obtained for arbitrary bare charges and salt content.

\begin{figure}[tbh]
\vskip 2mm
$$\epsfig{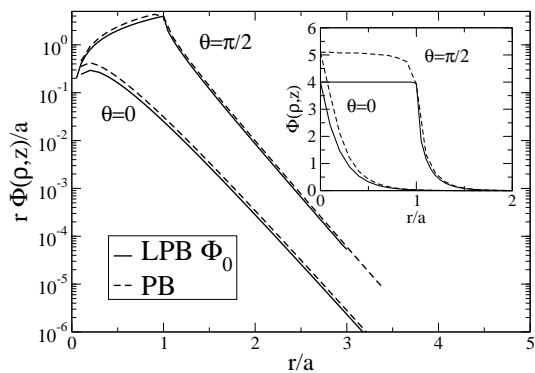}$$
\caption{\label{fig:fig9} Dashed line: electrostatic potential solution 
of PB equation (\ref{eq:pb}) for a highly charged disc with uniform bare
surface charge (Neumann-like boundary condition, 
$Z_{\bare}\,\lb/a=60$). Here $\kappa a=5$.
The continuous line shows the solution of LPB theory
with Dirichlet boundary condition $\phi_0=4$ (see section \ref{sec:fredholm}).
The behaviour is shown along the two perpendicular directions
$\theta=0$ and $\theta=\pi/2$. The main graph shows $r \phi$ 
versus $r/a=(\rho^2+z^2)^{1/2}/a$ on a linear-log scale. The inset shows
the previous potentials vs $r$ on a linear scale. }
\end{figure}

To test the constant potential picture put forward in
section \ref{sec:why} (which dwells on the fact that $\kappa a$ 
is ``large enough'') we show in Fig. \ref{fig:fig9} the PB
potential corresponding to a ``large'' bare charge.
It appears that the PB and LPB potential are in excellent agreement
except in the immediate vicinity of the disc, so that the constant effective 
potential prescription seems accurate. 

From the PB potentials, we may also extract the effective charge
$Z_{\eff}$ and anisotropy function $f(\kappa a,\theta)$,
that convey a more complete information than a plot like that
of Fig. \ref{fig:fig9}.
$Z_{\eff}$ follows from the far-field behaviour along the $\theta=0$ axis
\cite{Rque3}
[see Eq. (\ref{eq:normalaxis})]:
\be
\phi(r,0) \sim  Z_{\eff}\, \lb\, \frac{e^{-\kappa r}
}{r}.
\ee
Once $Z_{\eff}$ is known, $f$ is computed from Eq. (\ref{eq:ff}).

\begin{figure}[tbh]
\vskip 2mm
$$\epsfig{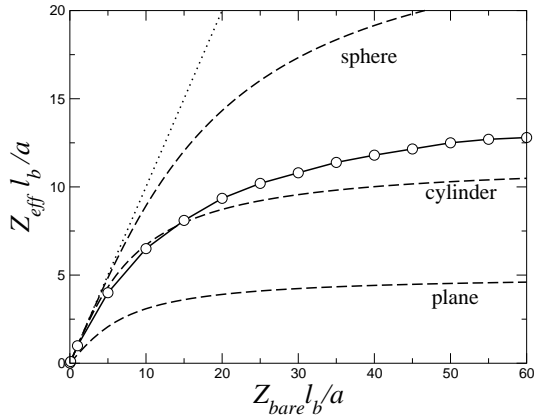}$$
\caption{\label{fig:fig10} Effective charge as a function
of the bare one, for $\kappa a=5$. The circles correspond
to the PB solution, and the dotted line has slope one to
show the regime of weak coupling where $Z_{\eff}=Z_{\bare}$.
The analytical expressions for a plane, a cylinder and a sphere
have also been plotted (see labels).}
\end{figure}

\begin{figure}[tbh]
\vskip 2mm
$$\epsfig{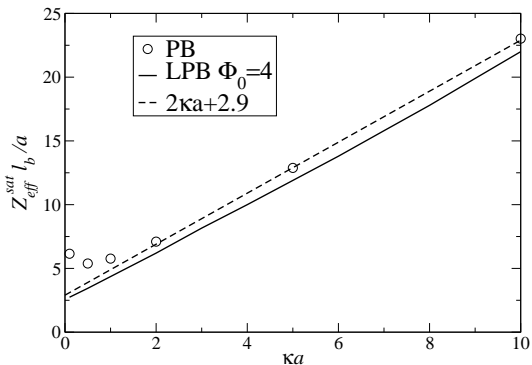}$$
\caption{\label{fig:fig11} Saturated effective charge $Z_{\eff}^{\sat}$
as a function of salt (circles). The LPB charge $Z$ is also shown
for $\phi_0=4$  (continuous line). The dashed line displays the 
empirical expression (\ref{eq:empirical}). }
\end{figure}

For fixed $\kappa a$, $Z_{\eff}$ is a function of the bare 
charge, see Fig. \ref{fig:fig10}, where the corresponding
analytical predictions for planes, spheres and cylinders (of
infinite length) have also been reported \cite{JPA}.
Not surprisingly, the behaviour is intermediate between
that of infinite planes and spheres, and somehow resembles
the results valid for charged cylinders. In addition,
$Z_{\eff}$ reaches a saturation plateau when $Z_{\bare}$ becomes
large \cite{Presc} (see Fig. \ref{fig:fig10}). 
This asymptotic plateau defines
the effective charge at saturation $Z_{\eff}^{\sat}$, 
which is shown in 
Fig. \ref{fig:fig11}. This quantity is an increasing function 
of salt content (except for $\kappa a<0.5$, see below), 
since an increase in salt density
enhances the macroion/microion screening, which diminishes
the amount of ``counterion condensation'' and consequently
increases the effective charge \cite{Presc}.
Figure \ref{fig:fig11} shows that the constant potential 
prescription of section \ref{sec:why} with $\phi_0=4$ 
provides a satisfying description of highly charged platelets,
as far as the (effective) charge is concerned.
It also appears that the best linear interpolation
reads
\be
Z_{\eff}^{\sat}\lb/a \simeq 2 \kappa a + 2.9
\label{eq:empirical}
\ee
which is rather close to approximation (\ref{eq:Zestime})
with the choice $\phi_0=4$. It may be observed in Fig.
\ref{fig:fig11} that for low $\kappa a$, the effective charge 
increases when $\kappa a$ decreases. This aspect will be 
discussed in section \ref{ssec:low}

\begin{figure}[tbh]
\vskip 2mm
$$\epsfig{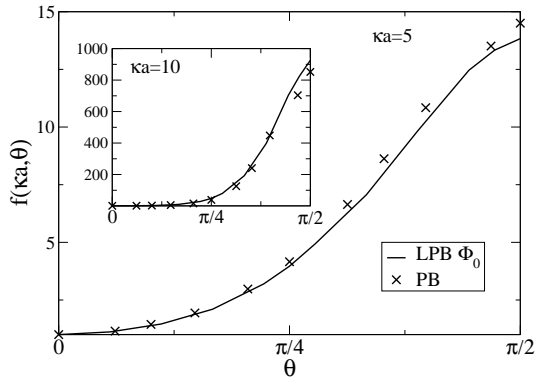}$$
\caption{\label{fig:fig12} Anisotropy function $f(\kappa a,\theta)$ 
as a function of azimuthal angle, for $\kappa a=5$. The crosses 
correspond to the PB result with $Z_{\bare} \,\lb /a=60$ (i.e. in
the saturation regime) while the continuous curve shows the constant 
potential LPB
result obtained in section \ref{sec:fredholm}. Inset: same for $\kappa a=10$.}
\end{figure}

\begin{figure}[tbh]
\vskip 2mm
$$\epsfig{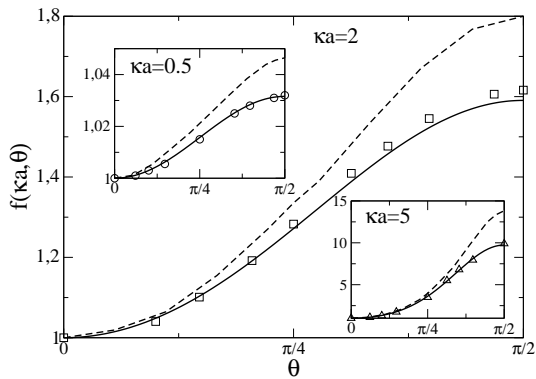}$$
\caption{\label{fig:fig13} Same as Fig. \ref{fig:fig12}, for a low bare
charge $Z_{\bare} \,\lb /a=10^{-2}$. 
The circles, squares and triangles correspond to
$\kappa a=0.5$, 2 and 5. 
The prediction of Eq. (\ref{eq:I1})
is shown by the continuous line. For the sake of comparison, we also plot 
with a dashed line the constant
potential LPB result already displayed in Fig. \ref{fig:fig12},
which is relevant at high bare charges.}
\end{figure}

Computation of anisotropy functions confirms the relevance of the
constant effective potential picture (see the comparison
proposed in Fig. \ref{fig:fig12}).
As shown in the inset, the very high values $f\simeq 900$ predicted 
from the analysis of section \ref{sec:fredholm} are indeed found
within non-linear PB. Note that the agreement reported in Fig. 
\ref{fig:fig12} is only expected at high bare charges. 
For low bare charges, $f(\kappa a,\theta)$ depends 
--at variance with its large bare charge counterpart --on the details
on the boundary conditions chosen on the disc to solve PB.
In the present situation (uniform surface charge), $f$ may be computed
analytically with the result \cite{JPCM}
\be
f(\kappa a,\theta) \,=\, 2\,\frac{I_1(\kappa a \sin\theta)}{\kappa a \sin\theta}.
\label{eq:I1}
\ee
This functional form is observed from our numerical data, for $Z_{\bare}\,\lb/a <1$
(see Fig. \ref{fig:fig13}). It turns out to differ much from 
that reported in section \ref{sec:fredholm} (shown with a 
dashed line in Fig. \ref{fig:fig13}). We may also observe 
in Fig. \ref{fig:fig13} that the 
``Neumann'' expression (\ref{eq:I1}) is lower than the Dirichlet one.
The reason is the following : $f$ is sensitive to the charges
lying near the edge of the disc [see e.g. 
Eqs. (\ref{eq:fsigma}) and  (\ref{eq:fI0})]. 
With Dirichlet boundary condition,
the induced surface charge diverges (see appendix \ref{app:C}),
at variance with the situation of a uniform surface charge,
which therefore exhibits a less anisotropic potential.
The Dirichlet and Neumann expressions respectively provide upper 
and lower bounds for the anisotropy function.

\begin{figure}[tbh]
\vskip 2mm
$$\epsfig{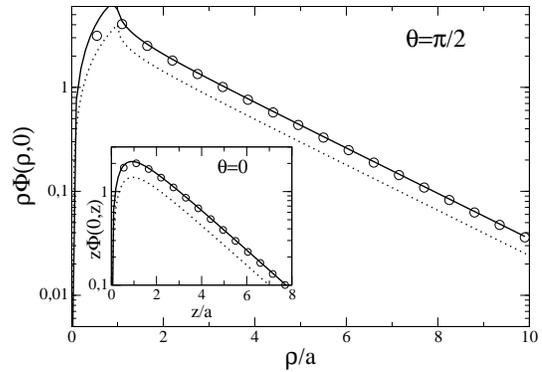}$$
\caption{\label{fig:fig14} Plot of the PB potential (continuous line)
versus distance from disc center in the $\theta=\pi/2$ direction
(disc plane). Also displayed are the LPB results for 
$\phi_0=\phi_0^{\hbox{\scriptsize opt}} = 5.7$
(circles) and for $\phi_0=4$ (dotted curve).
Here, $\kappa a=0.5$ and $Z_{\bare} \lb /a=15$
corresponding to the saturation plateau (the precise value of
$Z_{\bare}$ is therefore irrelevant).
The inset shows the same quantities along the disc normal.}
\end{figure}

\begin{figure}[tbh]
\vskip 2mm
$$\epsfig{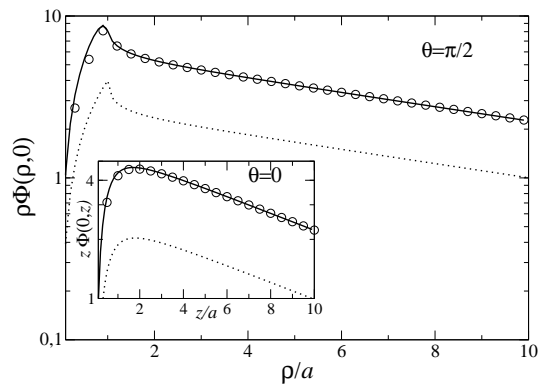}$$
\caption{\label{fig:fig14b} Same as Fig. \ref{fig:fig14} for 
$\kappa a =0.1$, with now $\phi_0=\phi_0^{\hbox{\scriptsize opt}}=9.0$.}
\end{figure}

\begin{figure}[tbh]
\vskip 2mm
$$\epsfig{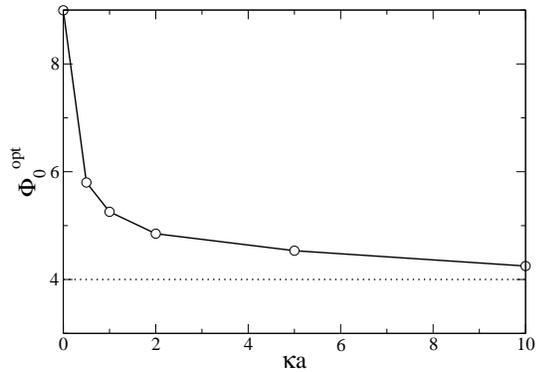}$$
\caption{\label{fig:fig15} Circles: Values of the optimal surface potential
$\phi_0^{\hbox{\scriptsize opt}}$
to be imposed within LPB theory to produce the same far-field
behaviour as a highly charged disc treated within PB (saturation limit).
The dashed line indicates the high salt limiting behaviour
$\phi_0=4$.}
\end{figure}

\begin{figure}[tbh]
\vskip 2mm
$$\epsfig{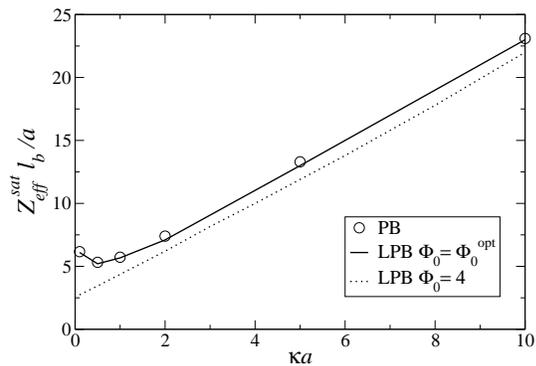}$$
\caption{\label{fig:fig16} Same as Fig. \ref{fig:fig11} but with 
$\phi_0=\phi_0^{\hbox{\scriptsize opt}}$ (the latter quantity being plotted 
in Fig. \ref{fig:fig15}) instead of $\phi_0=4$. For completeness,
the results obtained with $\phi_0=4$ are shown by the
dotted line.}
\end{figure}

In spite of the good agreement shown in Fig. \ref{fig:fig9},
a slight difference may be observed 
between PB and LPB results. It may be concluded
that a value $\phi_0$ slightly above 4 may give a better agreement
between non-linear and linear profiles. As mentioned at the
end of section \ref{sec:why}, finite $\kappa a$ corrections 
increase the value of the effective potential above 4 for
spherical and rod-like macroions. Figure \ref{fig:fig14}
and \ref{fig:fig14b}
show
that a similar effect exists for discs : when the $\phi_0$ 
of LPB approach is considered as an adjustable parameter,
the agreement between PB and LPB (again for highly charged discs)
becomes excellent even at relatively low values of $\kappa a$,
such as $\kappa a=0.5$ or even $\kappa a=0.1$.
The salt dependence of the above optimal potential,
denoted $\phi_0^{\hbox{\scriptsize opt}}$, is shown in 
Fig. \ref{fig:fig15}. 
One observes that $\phi_0^{\hbox{\scriptsize opt}}$ is
close to 4 for $\kappa a>5$, but may take significantly 
different values at lower $\kappa a$. This leads to reconsider the
plot of Fig. \ref{fig:fig11} since the (relative) discrepancy
PB/LPB may arise from discarding finite $\kappa a$ effects 
(i.e. enforcing $\phi_0=4$). Fig. \ref{fig:fig16} compares PB 
saturated effective charge to its LPB counterpart, with 
$\phi_0=\phi_0^{\hbox{\scriptsize opt}}$. Both quantities
now agree very well. This latter comparison is a severe 
and successful test for the relevance of the constant 
potential prescription.

\section{Discussion}
\label{sec:remarks}

\subsection{Pair potential}
From the properties of the one body electrostatic potential $\phi$
discussed previously, one may obtain the large distance behaviour
for the pair potential $U_{12}$ in the situation of two discs
(radii $a_1$ and $a_2$) in an electrolyte
\be
U_{12}\, = \,Z_{\hbox{\scriptsize eff,1}}\,
Z_{\hbox{\scriptsize eff,2}}\, \, \lb\, f(\kappa\, a_1, \theta_1)\,
f(\kappa\, a_2, \theta_2) \, \, \frac{e^{-\kappa r}}{r}.
\label{eq:U12}
\ee
Here $\theta_i$ is the angle between the normal to disc $i$ and the
center-to-center direction ${\bf r}_{12}$ (with $|{\bf r}_{12}|=r$). 
The validity of such an expression at intermediate or short
distances (i.e. $\kappa r$ of order 1) is unclear, since polarization
effects of disc $i$ on disc $j$ should at least perturb the symmetry
of the effective charge distribution, and hence alter the one
body expression for $f$ plotted in Fig. \ref{fig:fig12}.
We also note that the sub-leading terms in Eq. (\ref{eq:U12})
that become more important as $\kappa r$ decreases, involve
a more complex dependence on relative orientations (with all
Euler angles becoming relevant, contrary to the far-field
case where only $\theta_1$ and $\theta_2$ matter).

We may conclude here that at fixed center-to-center distance $r$, 
the favored configuration is that where the discs are parallel
and perpendicular to their center-to-center vector 
${\bf r}_{12}$. The {\tt T}-shape configuration is intermediate
and the most repulsive one corresponds to coplanar discs
(parallel to ${\bf r}_{12}$, as coins lying on a table).
However, the situation changes if one fixes the closest distance
$\cal D$
between the two discs. Comparing the configuration
$\theta_1=\theta_2=\pi/2$ where ${\cal D} = r-a_1-a_2$ 
to that with $\theta_1=\theta_2=0$ for which ${\cal D} = r$
requires to compare $\cal Q$ defined as
\be
{\cal Q} \,=\, f\left(\kappa\, a_1, \frac{\pi}{2}\right)\,
f\left(\kappa\, a_2, \frac{\pi}{2}\right)\, \, e^{-\kappa (a_1+a_2)}
\ee
with 1. From approximate expression (\ref{eq:I1}), 
it appears that $\cal Q$ is always smaller than 1. 
Instead of (\ref{eq:I1}), a more reliable expression for the
anisotropy is provided by (\ref{eq:f2param}),
which leads to the same conclusion. 
We therefore recover the intuitive result
that the less repulsive configuration 
at fixed $\cal D$ is for $\theta_1=\theta_2=\pi/2$
(two coins on a table).

\subsection{Behaviour at low $\kappa a$}
\label{ssec:low}

In the present study, we have focused on the regime
$\kappa a>1$ since according to the argument of
section \ref{sec:why}, it corresponds to the situation
where the effective potential may be predicted analytically.
It appears that the Debye length acts as a local probe to reveal 
the anisotropy of the macroion under study. Hence, in the limit
of small $\kappa a$ where this probe cannot resolve
the disc dimension, we found that the anisotropy disappears: 
$f(\kappa a=0,\theta)=1,~ \forall \,\theta$. We may then speculate
that at small $\kappa a$, the precise form of the macroion
becomes irrelevant so that we should recover the same results
as for spheres \cite{Rque5}. From the analysis of Ramanathan \cite{Ramanathan},
we may consequently expect in the saturation regime~:
\be
\Zeff^{\sat}\,\frac{\lb}{a} ~ \stackrel{\kappa a \ll 1}{\sim}  ~
 -2\ln(\kappa a) + 2 \ln[-\ln(\kappa a)] + 4 \ln 2 
\label{eq:Raman}
\ee
Such an expression diverges for $\kappa a \to 0$,
indicating that potential (or charge) renormalization 
ultimately becomes irrelevant. However, with the lowest
value of $\kappa a$ investigated in this work
($\kappa a=0.1$), we have measured $\Zeff^{\sat} \simeq 6.1 \,a/\lb$
(see Fig. \ref{fig:fig11}),
whereas Eq. (\ref{eq:Raman}) gives approximately 
a value 6.7. Whether this agreement is incidental or not is unclear.
We also note that if the discs behave as spheres for 
very low $\kappa a$, their saturated effective potential 
should coincide with $\Zeff^{\sat} \,\lb/a$. This is not the 
case for $\kappa a=0.1$ where we measured
$\phi_0^{\hbox{\scriptsize opt}}\simeq 9$.
This may indicate what a salinity condition $\kappa a=1/10$ 
is not low enough to enter the ``spherical'' regime.

\subsection{Validity of the PB approach} 

We now discuss the validity of the Poisson-Boltzmann theory
underlying the present analysis. Such an approach neglects
microionic correlations (be they of electrostatic or other
origin, such as excluded volume) while macroion/microion
electrostatic correlations are correctly incorporated.
In the vicinity of the charged discs where the counterion
density may become large, the neglected correlations 
are most important, and may invalidate PB theory
if the disc bare charge is too large (say 
$\Zbare>\Zbare^{\hbox{\scriptsize corr}}$). 
Since we have considered
here the situation of high $\Zbare$ to explore the PB saturation 
plateau, we need to justify the relevance of such a plateau.
In other words, this amounts to elucidating the circumstances
under which $\Zsat < \Zbare^{\hbox{\scriptsize corr}}$,
since for $\Zbare > \Zsat$, one has, roughly speaking,
$\Zeff \simeq \Zsat$.

\begin{figure}[tbh]
\vskip 2mm
$$\epsfig{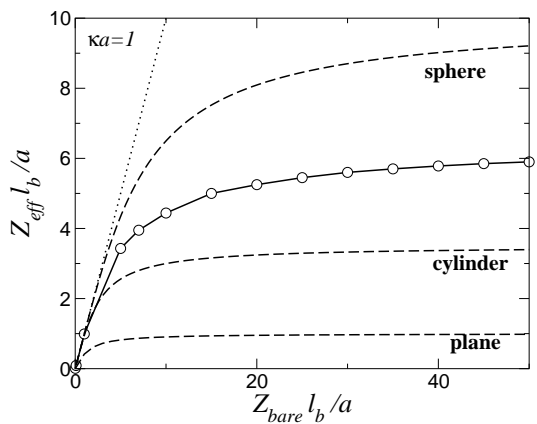}$$
\caption{\label{fig:fig17} Same as Fig. \ref{fig:fig10} with $\kappa a=1$.}
\end{figure}

For a salt-free system, Netz has considered the validity of
PB theory in planar, cylindrical and spherical geometries
\cite{Netz}. Since no general result exists, we present here a
a simple argument concerning
discs, which goes as follows (see \cite{Levin} for the spherical case). 
Microionic correlations may be
accounted for by the coupling parameter $\Gamma = \lb/d$
where $d$ is a typical distance between microions in the 
double-layer. This distance is bounded from below by that,
denoted $d^*$, 
where all $Z$ monovalent counterions are artificially condensed onto
the disc (as would happen in the low temperature limit).
It may be estimated writing that the typical surface per
microion $(d^*)^2$ on the disc is the mean value $\pi a^2/Z$.
Hence, 
\be
\Gamma\, \simeq \,\sqrt{\lb^2\,\sigma_{\bare}} \simeq 
\sqrt{\Zbare \,\lb^2/(\pi a^2)}.
\ee
For microions with valency $z$, one would obtain
\be
\Gamma \simeq \sqrt{z^3\, \Zbare \,\lb^2/(\pi a^2)}.
\label{eq:Gammaz}
\ee
For the sake of the argument, the factor $\pi$ could be omitted.
The important point here is that $\Zbare\, \lb/a$ may be large,
which corresponds to the saturation regime of PB theory,
with still $\Gamma<1$, which justifies the mean-field assumption
underlying PB. With typical Laponite parameters \cite{Rque1}
and monovalent microions, we have $\Gamma\simeq 0.7$ for 
$\Zbare \simeq 700$, a reasonable value for the charge. In addition,
$\Zbare\, \lb/a \simeq 33$ which is well beyond the linear regime
where effective and bare parameters coincide (see Fig. \ref{fig:fig10},
or Fig. \ref{fig:fig17} corresponding to a lower salt concentration
for which $\Zbare\, \lb/a \simeq 33$ lies in the saturated region).
We finally note that the intersection between the PB saturation
regime and the consistency condition $\Gamma<1$ is all the larger
as $a/\lb$ is big \cite{Groot}.  
But of course, when $\Zbare$ strictly diverges,
$\Gamma$ exceeds a few units and PB breaks down. Finally, to be
specific, $\Zbare^{\hbox{\scriptsize corr}}$ would be defined 
as the value of $\Zbare$ such that the coupling parameter 
$\Gamma$ is of order unity. The case of Laponite seems somehow
borderline, since $\Zbare$ does not
differ much from $\Zbare^{\hbox{\scriptsize corr}}$
(on the order of a few thousands).

\subsection{The case of asymmetric electrolytes}

Bearing in mind the classical rule that increasing the valency 
of microions decreases the range of validity of PB theory
[see Eq. (\ref{eq:Gammaz})], we may extend 
previous results to 1:2 and 2:1 electrolytes. The key ingredient
in the constant effective potential prescription is indeed
the analytical solution of the planar (1D) PB equation. 
The latter problem has been solved by Gouy almost a
century ago \cite{Gouy}, and it turns out that the counterpart
of the monovalent result $\phi_0=4$ reads
$\phi_0 \,=\,  6 $
for 2:1 electrolytes (i.e monovalent counterions and divalent
coions, or more precisely when the ratio of coion to counterion
valency equals 2). In the reverse 1:2 situation, we have 
\be
\phi_0 \,=\, 6 (2-\sqrt{3}) \simeq 1.608.
\ee
The 1:2 effective potential is smaller that the 2:1 potential since 
screening by monovalent counterions is less efficient than
with divalent ones (hence a higher effective potential, and a higher
effective charge \cite{Tellez}). 

By simply plugging the above expressions for $\phi_0$ into the 
expressions derived in the previous sections for symmetric electrolytes,
one may describe 1:2 and 2:1 situations as well. We finally note
that the electrolyte asymmetry does not affect the anisotropy function
$f(\kappa a,\theta)$: LPB equation takes the same form 
(modulo a change in the numerical value of $\kappa$)
and only $\phi_0$ is affected.

\subsection{Comparison with existing results}

In Ref \cite{JPCM}, we have addressed a similar issue as in the present
paper. However, 
neither the LPB at constant potential nor the PB theory
were  solved. The anisotropy function has been estimated there
from the Neumann LPB result with a uniform surface charge.
This leads to expression (\ref{eq:I1}), which has been 
shown in Figure \ref{fig:fig13} be an underestimation 
of the Dirichlet result [and the agreement between PB and 
Eq. (\ref{eq:I1}) is simply due to the low charge
in Fig. \ref{fig:fig13}, see Fig. \ref{fig:fig12}].

Following similar lines, the saturated effective charge of
discs have been estimated in \cite{JPCM} from the Neumann
LPB solution. When the resulting dimensionless 
potential is equated to 4 on the disc
center, we obtain \cite{JPCM}
\be
\Zeff^{\sat} \,=\, \frac{a}{\lb} \, \frac{2\, \kappa a}{1-\exp(-\kappa a)},
\ee
which is very close to $2\kappa a$ as soon as $\kappa a>2$. Such an expression
only captures the leading order behaviour (the planar limit), but misses
the offset correction (2.9) as appears in Eq. (\ref{eq:empirical}).

\section{Conclusion}
\label{sec:concl}

We have presented in this paper a detailed comparison between
the electrostatic potentials obtained within Poisson-Boltzmann (PB) and 
Linearized PB approximations, for a charged disc in an electrolyte.
We have proposed a new and efficient semi-analytical 
method to solve the LPB problem at constant surface potential $\phi_0$.
The procedure used is not restricted to the specific problem 
considered here, and allows to solve more general  situations of the 
form given by Eq. (\ref{eq:S1plusku}).
On the other hand, the PB problem has been solved numerically following
similar lines as in Ref \cite{Raul}. We have shown that the 
far-field potential created by a highly charged disc within
PB is remarkably close to its LPB counterpart with a suitably
chosen value of $\phi_0$, which therefore defines
the effective potential of the charged discs. 
As expected from the argument
put forward in section \ref{sec:why}, the latter quantity 
is close to 4 (meaning that the effective surface potential
is close to $4 kT/e$) whenever $\kappa a$ is larger than a few units,
say $\kappa a> 3$. These results extend the conclusion of Ref.
\cite{Presc}. The argument of section \ref{sec:why} should in fact 
apply for any charged macro-ion for which the curvature is smaller
than the inverse Debye length $\kappa$ of the surrounding electrolyte.
Note also that in the limit of high bare charge, 
the details of the bare charge distribution onto the
discs are irrelevant. In this respect, the results obtained 
here within PB with uniform surface charge are generic
and would resist to charge modulation.

The scenario emerging is that due to non-linear screening phenomena, 
highly charged macroions may be considered as effective
objects with a uniform surface potential and can be treated
within a linear theory (provided short distance features are irrelevant)
which considerably simplifies the analysis and opens simulation
routes. 
In addition, this potential
is constant provided there is enough salt in the solution,
in the sense that it no longer depends on physico-chemical
parameters. Such a viewpoint not only predicts satisfactorily
their effective charge, but also reproduces accurately the
anisotropy of their potential. The latter property, embodied in the function
$f(\kappa a,\theta)$ is a key feature of screened electrostatic
interactions and may have non negligible --and hitherto largely unexplored--
consequences. It is responsible for the rich phase behaviour 
and orientational ordering of colloidal molecular crystals
\cite{PRL}. Its effects on the phase behaviour of clays, especially
at moderately to high salt concentrations where large
energy barriers $f(\theta=\pi/2)-f(\theta=0)$ are observed, will be the 
subject of future work.

\vskip 2mm

The authors acknowledge fruitful discussions with J.J. Weis, B. Jancovici,
F. van Wijland, M. Aubouy and H. Lekkerkerker.

\begin{appendix}
\section{}
\label{app:A}

Mixed boundary value problems are rather frequent in electrostatics but also in 
diffusion and elasticity problems \cite{Lowengrub,Collins} (conduction of 
heat, diffusion of thermal neutrons, punch or crack problems etc.).
They may be encountered in hydrodynamics as well \cite{Lamb}. 
They generally arise whenever a potential
is prescribed over part of a boundary whereas its normal derivative
is specified over the complementary part. The
theory of dual integral equations turns out to be a powerful tool 
for such situations. In this appendix, we give more details about the problem
of finding the solution $A(k)$ to equations (\ref{eq:fa}) and (\ref{eq:fb}).
To begin with, it is convenient to recast them in the form
\be
\left\{	\begin{array}{lcrr}
	\displaystyle \int_{0}^{\infty} \frac{\displaystyle g(u)}{\sqrt{u^2+(\kappa a)^2}}
		\,J_{0}(x u)\,du &=& 1 & \quad \hbox{for} \,\, x < 1 \\
	\displaystyle\int_{0}^{\infty} g(u)\,J_{0}(x u)\,du &=& 0 & \quad \hbox{for} \,\, x > 1 
	\end{array} \right.
\label{eq:A-mixbound}
\ee
where dimensionless quantities have been introduced: $u=k\,a$, $x=\rho/a$ and 
$g(u)=\sqrt{u^2+(\kappa a)^2}\,A(u)/\Phi_{0}$. 
Solutions of the previous equations for $\kappa=0$ (no salt case) 
have been derived by Titchmarsh \cite{Titchmarsh}. The
procedure, based on rephrasing
the dual integral equations by means of some invertible linear operators 
gives the solution
\be
A_{\kappa=0}(u)\,=\,\frac{2\Phi_{0}}{\pi}\,\frac{\sin u}{u}
\label{eq:caslimit}
\ee
and  the corresponding potential is that which leads to equation 
(\ref{eq:sigmadiverge})
for the charge.
A generalization of Titchmarsh's method has been
proposed by Sneddon \cite{Sneddon} for dual integral equations of the type
\be
	\left\{ \begin{array}{lcrr}
	\displaystyle\int_{0}^{\infty} u^{-2\alpha}\,(1+\omega(u))\,
	g(u)\, J_{\nu}(x u)\,du &=& 
		1 & \quad \hbox{for}\,\,x < 1 \\
	\displaystyle\int_{0}^{\infty} g(u) J_{\nu}(x u)\,du &=& 
		0 &  \quad \hbox{  for}\,\,x > 1    
	\end{array}  \right.
\label{eq:S1plusku}
\ee
where $\omega$ is an arbitrary function. 
 Equations 
(\ref{eq:S1plusku}) and (\ref{eq:A-mixbound}) can be made equivalent by taking 
\be
	\omega(u)\,=\,\frac{u}{\sqrt{u^2+(\kappa a)^2}} - 1
\ee
with $\alpha=1/2$ and $\nu=0$ (for $\omega(u)=0$, we recover the no salt case). 
The problem at hand --that fits into the general framework of \cite{Sneddon}--
may be reduced
to that of solving a Fredholm equation of the second kind. Following
Sneddon, we write equations (\ref{eq:S1plusku}) in the form:
\be
	 \begin{array}{lcl}
		\hbox{S}_{-\frac{1}{2},1}[(1+\omega(u)) g(u)/u,x] &=& 2/x 
		\quad \hbox{for}\,\,x<1\\
		\hbox{S}_{0,0}[g(u)/u,x] &=& 0 \quad \hbox{    for}\,\,x>1
		\end{array} 
\label{eq:ku}
\ee
where $\hbox{S}_{\alpha,\beta}$ is the modified Hankel operator defined by:
\begin{eqnarray*}
\hbox{S}_{\alpha,\beta}[\lambda(u),x] &\equiv &
\hbox{S}_{\alpha,\beta}\lambda(x)\\
&=& 2^{\beta}\,
	x^{-\beta}\,\int_{0}^{\infty}u^{-\beta}\,\lambda (u)\, 
	\hbox{J}_{2\alpha+\beta}(xu)\,du
\label{eq:opera}
\end{eqnarray*}
We then introduce the function $h$ through
\be
g(u)\,=\,u \,S_{0,\frac{1}{2}}h(u).
\label{eq:psire}
\ee
This function $h(u)$ is the
central object in the present procedure.
After some cumbersome algebra based on the properties of the modified 
Hankel operators (for details, see \cite{Sneddon}), we find that the function 
$h_{2}=u\,h(u)$ defined on $[1;\infty]$ vanishes while its counterpart 
$h_{1}$ defined on $[0;1]$ 
is the solution of the following Fredholm equation of the second kind:
\be
h_{1}(x)+\int_{0}^{1}h_{1}(u)K(x,u)du\,=\,2/\sqrt{\pi}.
\label{eq:Fred}
\ee
The kernel $K(x,u)$ is defined in the whole $xu$-plane by:
\be
K(x,u)\,=\,\sqrt{\frac{\pi}{2}}\kappa a(I_{1}(\kappa a|x-u|)-L_{1}(\kappa a(x+u)))
\ee
and $I_{1}$ and $L_{1}$ denote respectively the modified Bessel function and Struve
function of the first kind, of order one.
The weight function $g(u)$ appearing in equation (\ref{eq:A-mixbound}) 
is recovered by means of the integral:
\begin{eqnarray}
g(u)&=&\sqrt{u^2+(\kappa a)^2}\,A(u)/\Phi_{0}\\
&=&\frac{u}{\sqrt{\pi}}\,\int_{0}^{1}\cos{(ut)}\,h_{1}(t)\,dt.
\label{eq:fubis}
\end{eqnarray}
The potential finally follows from equation (\ref{eq:sollpb}).
The Fredholm equation (\ref{eq:Fred}) is solved numerically by an 
iterative procedure, starting 
by an constant initial guess for $h_{1}$. 

\begin{center}
\begin{figure}[h]
\vskip 2mm
\epsfig{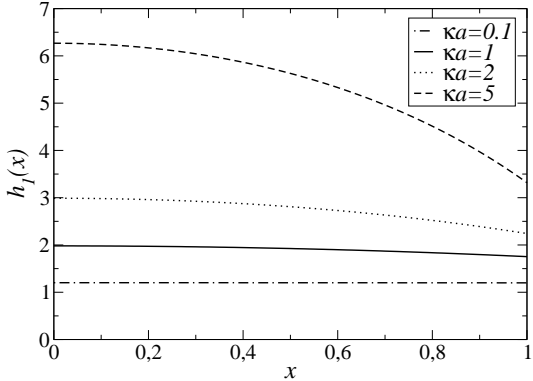}
\caption{Plots of $h_1(x)$ [solution of the integral equation (\ref{eq:Fred})] 
versus rescaled distance 
$x$ for different values of $\kappa a$. For $\kappa a=0.1$, $h_1$ is
very close to its no salt limit $2/\sqrt{\pi} \simeq 1.13$. }
\label{fig:h11}
\end{figure}
\end{center}

In the limit $\kappa=0$, $K(x,u)=0$ so that, from Eq. (\ref{eq:Fred}), $h_1=2/\sqrt{\pi}$. 
The function $g(u)$ in (\ref{eq:fubis}) follows
immediately: $g(u)\,=\,(2/\pi) \sin u$, which is fully consistent with
(\ref{eq:caslimit}).
Finally note that the functions $h_{1}(x)$, plotted in Fig.
\ref{fig:h11} for different values of
$\kappa a$, are related to the surface charge density of the disk $\sigma(x)$
through the equation:
\be
\sigma(x)\,=\,\frac{h_{1}(x)}{\sqrt{\pi}\,\sqrt{1-x^2}}.
\label{sh}
\ee
The function $h_1$ can be well approximated by a quadratic polynomial.
The corresponding charge $\sigma(x)$ may then be used to 
compute analytically approximated effective charges,
anisotropy functions and weigh functions $A(u)$. This procedure is an
alternative to the ansatz proposed in equation (\ref{eq:sig2param}),
the latter being more suited for an analytical treatment.
We finally note that from Eq. (\ref{sh}) and the regular behaviour of
$h_1$ observed on Fig. \ref{fig:h11} for $x\to 1$, the surface charge
diverges near the edge of the disc like $(1-x)^{-1/2}$, see appendix
\ref{app:C}.


\section{}
\label{app:C}

We show here that the LPB surface charge distribution $\sigma(\rho)$ 
--arising from the condition of constant surface potential on the disc--
exhibits in an electrolyte the same edge effect as in vacuum,
where it diverges as $(a-\rho)^{-1/2}$ when $\rho\to a^-$ \cite{Landau}.

\begin{figure}[htb]
$$\epsfig{figure=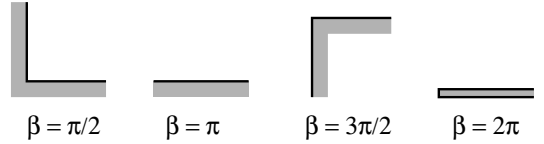,width=7.cm,angle=0}$$
\caption{\label{fig:fig20} The wedge geometry. The conductor is represented
as the shadowed part and its boundaries by the thick line.}
\end{figure}

We consider the more general problem of a wedged-shape conductor with
angle $\beta$ (see Fig. \ref{fig:fig20}). 
The imposed potential is denoted $\phi_0$. 
Being interested in the behaviour near the sharp edge
where the situation is of cylindrical symmetry, 
we introduce the cylindrical coordinates ($r$, $\psi$) in a plane
perpendicular to the apex of the wedge. 
With rescaled distance $\widetilde r = \kappa r$, we look for
solutions of LPB equation (\ref{eq:lpb}) 
\be
\widetilde r \, \frac{\partial}{\partial \widetilde r}\left(
\widetilde r \,\frac{\partial \phi}{\partial \widetilde r}\right) 
\,+\, \frac{\partial^2 \phi}{\partial \psi^2}
\,=\,\widetilde r^2\phi
\ee
in a form with separated
variables 
\be
\phi(\widetilde r,\psi) \,=\, R(\widetilde r) \,\Psi(\psi).
\ee

With the boundary condition 
$\phi(\widetilde r,\psi=0)=\phi(\widetilde r,\psi=\beta)=\phi_0$
for all $\widetilde r$ in the vicinity of the wedge (i.e. $\widetilde r$
close to 0), the solution reads:
\begin{eqnarray}
&&R(\widetilde r) = A_\nu I_\nu(\widetilde r) \\
&&\Psi(\psi) = \alpha_\nu  \sin(\nu \psi).
\end{eqnarray}
Here, $\nu= n \pi /\beta$ where $n\in\mathbb{N}$; $A_\nu$ and $\alpha_\nu$
are arbitrary constants. In the vicinity of the wedge, the potential therefore
takes the form
\be
\phi(\widetilde r,\psi) \,=\, \phi_0 \,+\,\sum_{n=1}^\infty A_n \sin\left(
\frac{n \pi \psi}{\beta} \right)\,I_{n \pi/\beta}( \widetilde r).
\ee
The dominant term when $\widetilde r \to 0$ corresponds to $n=1$,
and scales like $\widetilde r^{\,\pi/\beta}$. The associated surface
charge behaves like $\sigma(\widetilde r)\propto \widetilde r^{\,\pi/\beta-1}$,
hence like $\widetilde r^{\,-1/2}$ near the edge of a disc $(\beta=2 \pi)$.
This result has been used to choose the functional form (\ref{eq:sig2param}).

\end{appendix}



\begin{thebibliography}{90}
\bibitem{Rque1}
These colloidal discs bear a negative bare charge. They
have a radius of approximately 15 nm and a thickness
smaller than 1nm, see e.g.
Laponite technical report, Southern Clay Products Inc., Laporte plc.
The results reported here correspond to a positive bare charge.
They immediately transpose to the Laponite case by a change of sign.
\bibitem{Rque3}
The existence of a given direction along which the effective charge may be
directly measured is a specific feature of macroions with vanishing
internal volume. In general, as happens for a disc of finite thickness, 
the anisotropy function $f$ differs from
unity in all directions. It is consequently more difficult --and to some 
extent arbitrary-- to disentangle
effective charge effects from those due to anisotropy.
\bibitem{Rque5}
We are interested here again in the far-field behaviour.
For $\kappa =0$ (no electrolyte), a disc and a sphere
behave identically, as a usual monopole.
\end{thebibliography}
\end{document}